\documentclass[preprint,12pt]{elsarticle}

\usepackage{amsmath,amsfonts,amssymb}
\usepackage{graphicx}
\usepackage[colorlinks=true, allcolors=blue]{hyperref}

\journal{Analytica Chimica Acta}

\begin{document}

\begin{frontmatter}

\title{Bin Latent Transformer (BiLT): A shift-invariant autoencoder for calibration-free spectral unmixing of turbid media}

\author{Martin Hohmann}
\ead{Martin.Hohmann@FAU.de}

\address{Institute of Photonic Technologies (LPT), Friedrich-Alexander-Universit\"at Erlangen-N\"urnberg (FAU), Konrad-Zuse-Stra\ss e~3/5, 91052 Erlangen, Germany}

\begin{abstract}
The accurate recovery of constituent-level optical properties from integrating sphere measurements is a central analytical challenge in pharmaceutical analysis, food science, and biomedical diagnostics. Neural network autoencoders can extract spectrally resolved absorption and scattering coefficients for each constituent without prior knowledge, but their fully connected encoders bind learned features to absolute wavelength indices, causing accuracy loss under spectrometer calibration drift or hardware exchange. This work introduces the Bin Latent Transformer (BiLT)-Autoencoder, in which the dense encoder is replaced by a cross-attention scanner: 16 learnable probe vectors query a convolutional feature map, aggregating morphological spectral information independently of absolute wavelength position. A physics-constrained linear decoder with enforced absorption/scattering separation and a three-phase curriculum augmentation strategy complete the architecture. On a liquid phantom benchmark (intralipid and two ink absorbers; 496 samples), the model achieves $R^2 = 0.979$ and $0.975$ for $\mu_a(\lambda)$ and $\mu_s'(\lambda)$, respectively, on held-out test spectra, maintaining $R^2 > 0.90$ for $\mu_a$ and $R^2 \approx 0.99$ for $\mu_s'$ across the full tested shift range of $\pm 10$ spectral bands. The model generalises to a simulated spectrometer with a broader instrument line shape (${\approx}24$\,nm FWHM) without retraining, retaining $R^2 \approx 0.96$ and $0.974$ for the two channels. Attention map analysis reveals a physically interpretable two-component probe strategy: sparse anchor probes at absorption-edge wavelengths combined with a diffuse, SNR-driven ensemble at the high-transmittance long-wavelength region, which recruits additional probes dynamically under noise to provide implicit spectral averaging.
\end{abstract}


\end{frontmatter}

\section{Introduction}

The non-destructive quantitative characterisation of turbid, optically scattering materials is a central analytical challenge in pharmaceutical development, food quality control, and biomedical diagnostics~\cite{Messung_OP}. In all these domains, the measured diffuse reflectance spectrum represents a non-linear, entangled superposition of the absorption and scattering contributions of every constituent present, making direct quantitative interpretation intractable without a physical model of light transport. The standard approach recovers the spectrally resolved absorption coefficient~$\mu_a(\lambda)$ and reduced scattering coefficient~$\mu_s'(\lambda)$, collectively called the optical properties~(OP), from total reflectance and transmittance measurements performed with an integrating sphere, followed by inverse Monte Carlo simulations~(IMCS)~\cite{integration_sphere_precise1,integration_sphere_precise2}. While physically rigorous, this pipeline is computationally intensive and delivers only bulk, constituent-averaged OP: aggregate quantities that reflect the combined effect of all constituents without identifying or quantifying any of them individually.

Machine learning approaches, and neural networks~(NNs) in particular, have emerged as practical alternatives to IMCS for inverting integrating sphere measurements. The approach was pioneered by Farrell et al.~\cite{farrell1992use}, who demonstrated that a NN trained on spatially resolved diffuse reflectance measurements could recover $\mu_a$ and $\mu_s'$ without explicit analytical inversion. Subsequent work extended this to a variety of measurement geometries and configurations: the sub-diffusive regime of spatially resolved reflectance~\cite{ivancivc2018efficient}, broad-band double integrating sphere setups~\cite{hokr2021machine,nishimura2021determination}, thin tissue configurations~\cite{chen2023real}, and integrating sphere measurements that omit the collimated transmission channel~\cite{ni2024reconstruction}. In all cases, NN regressors achieve prediction speeds several orders of magnitude faster than IMCS at comparable accuracy. Despite this operational advance, these models share the fundamental limitation of the conventional OP pipeline: they output aggregate OP without resolving the contributions of the individual chemical constituents that generate them. The spectral fingerprint, identity, and relative concentration of each absorber and scatterer in the mixture remain inaccessible.

Recovering constituent-level information requires moving beyond bulk OP estimation into the framework of spectral unmixing: the decomposition of an observed spectrum into pure-component spectra~(endmembers) and their respective abundances. Neural network autoencoders offer a fully data-driven route to this decomposition without requiring the endmembers to be specified a priori~\cite{palsson2018hyperspectral,hong2021interpretable}. Physics-constrained variants that enforce non-negativity and abundance-sum constraints have demonstrated substantially improved interpretability and quantitative accuracy in hyperspectral and Raman imaging contexts~\cite{georgiev2024hyperspectral}. Building on this paradigm, Ni~et~al.\ introduced a mixed autoencoder architecture for optical property reconstruction that automatically recovers the spectral OP of each absorber and scatterer from integrating sphere measurements without any prior knowledge of the constituents~\cite{ni24,ni2025automated}.

A fundamental barrier to deploying such models beyond the laboratory has, however, remained unaddressed: their sensitivity to instrument-specific spectral characteristics. A dense~(fully connected) encoder explicitly binds each learned spectral feature to the absolute pixel index of the input. Any systematic wavelength offset (caused by spectrometer calibration drift, thermally induced grating shift, or use of a different instrument) simultaneously alters the input to every neuron, destroying the learned spectral representation. This sensitivity to absolute spectral position is a well-recognised challenge in analytical spectroscopy, broadly subsumed under the term calibration transfer~\cite{chatzidakis2019calibration}. Conventional remedies require additional reference measurements or post-hoc transfer algorithms; an architecture that is \textit{intrinsically} robust to spectral shifts would represent a qualitative step towards instrument-independent analytical spectroscopy.

The attention mechanism of the Transformer architecture~\cite{vaswani2017attention} offers a principled solution to this problem. Unlike dense layers that bind learned features to fixed input coordinates, attention computes pairwise interactions between elements of the input sequence, making the representation sensitive to the \textit{relative} structure of the spectrum (peak shapes, bandwidths, and inter-feature distances) while relaxing the rigid dependency on exact wavelength indices. Koyun et al.\ demonstrated this principle with the RamanFormer~\cite{koyun2024ramanformer}, an attention-based model that quantifies component concentrations in Raman mixture spectra with state-of-the-art accuracy, outperforming classical least-squares regression and established convolutional architectures. However, its self-attention operates on fixed spectral patches: a systematic wavelength offset shifts features across patch boundaries, reintroducing positional sensitivity equivalent to that of a dense encoder. Moreover, Raman bands are typically narrower than the spectrally broad absorption and scattering profiles encountered in integrating sphere measurements. The Byte Latent Transformer~(BLT)~\cite{pagnoni2024byte}, originally designed for text processing, avoids this limitation through a set of fixed latent probe vectors that query the input sequence via cross-attention. By combining local morphological feature extraction with macroscopic positional embeddings, the internal representation acts as a flexible searchlight scanning a regional spectral neighbourhood rather than being hard-coded to absolute positions.

In this work, the \textbf{Bin Latent Transformer (BiLT)}-Autoencoder is introduced: a shift-invariant, interpretable neural network framework for the simultaneous extraction of constituent-level absorption and scattering spectra from integrating sphere measurements of turbid media. The architecture builds on both the physics-constrained mixed autoencoder of Ni~et~al.~\cite{ni2025automated} and the latent probe mechanism of the BLT: the dense encoder of the mixed autoencoder is replaced by the BiLT scanner, preceded by a large-kernel convolutional feature extractor. The network is trained via a three-phase curriculum augmentation strategy that progressively introduces spectral shifts and additive noise, preventing catastrophic forgetting of the underlying physics. The physics-constrained linear decoder is retained in full. The resulting model maintains high predictive accuracy for both $\mu_a(\lambda)$ and $\mu_s'(\lambda)$ under severe perturbations~($\pm10$ spectral bands, up to 5\,\% additive noise). Furthermore, the attention maps of the transformer provide direct visualisation of which spectral regions the model identifies as analytically informative, providing a degree of interpretability not available in dense architectures.

\section{Methods}

\subsection{Experimental parameters}

The experimental parameters are identical to those of the preceding study~\cite{ni24} and the measurement procedure is described in full detail in~\cite{ni2024reconstruction}. For convenience, a brief summary is provided here. Total transmission~(TT) and total reflection~(TR) were recorded in the spectral range of 500--800~nm at a step size of 2~nm, yielding 150 spectral data points per channel, using a Shimadzu UV-3600 UV-VIS-NIR spectrophotometer~(double beam, three-detector system) equipped with the LISR-3100 large integrating sphere attachment. The liquid phantoms consisted of intralipid~(IL) as a scatterer and either red (Modena Red, MONTBLANC) or black (Indian Ink, Winsor \& Newton) ink as absorbers, prepared at a total of 496 concentration combinations. The spectrally resolved absorption coefficient~$\mu_a(\lambda)$ of each ink was determined from direct spectrophotometric measurements on ink solutions without scatterer; the reduced scattering coefficient~$\mu_s'(\lambda)$ of intralipid was taken from the literature~\cite{aernouts2014dependent}. These constituent-level spectra, scaled by concentration, serve as the ground-truth output for model training. In the present study, only measurements from the 5~mm cuvette are used as input. The data are divided into a training set~(80\%) and a test set~(20\%) by a random split. All input spectra are normalised to the range $[0, 1]$ prior to training.

\subsection{Machine learning}

\subsubsection{Network architecture}

\textbf{BiLT scanner (encoder).} The shift invariance of the BiLT scanner arises from a three-stage mechanism in which each processing step transforms the spatial properties of the representation in a specific way: (i)~translation-equivariant feature extraction, (ii)~local shift tolerance, and (iii)~spatial aggregation via cross-attention.

In the first stage, a one-dimensional convolutional layer (128~filters, kernel width~23, GELU activation, same padding) followed by layer normalisation (Layers~1--2, Table~\ref{tab_layers_nn}) extracts local spectral features such as peak shapes, shoulders, and gradients. Because the convolution applies identical learned filters at every spectral position, the resulting feature map is \emph{translation-equivariant}: if a spectral feature shifts by $k$~positions in the input, its activation in the feature map shifts by exactly $k$~positions as well, but its shape and magnitude remain unchanged. In the second stage, a max-pooling layer with pool size~3 and stride~1 (same padding, Layer~3) introduces additional local shift tolerance by selecting the maximum activation within a three-bin window at each position. This blurs the precise feature locations by approximately $\pm1$~bin without changing the sequence length, making the subsequent attention mechanism robust to sub-pixel misalignments. A second layer normalisation (Layer~4) stabilises the feature magnitudes before the positional encoding.

An importance gating mechanism (Layer~5) is applied to the normalised feature vectors before the positional encoding. Conceptually similar to the entropy-based scaling in the original BLT architecture~\cite{pagnoni2024byte}, this gate computes a scalar importance weight for each spectral position. At every position, three summary statistics are extracted across the 128~feature dimensions: the mean, the standard deviation, and the maximum activation. These three values are passed through a small multi-layer perceptron ($3 \rightarrow 16$ with ReLU, $16 \rightarrow 1$ with sigmoid), producing a gate score between~0 and~1 for each position. The feature vectors are then multiplied element-wise by this score, so that positions with informative spectral content (high variance, strong activations) are amplified while featureless or noisy regions are suppressed. Because the importance gating computes its scores independently at each spectral position and is applied before the positional embedding, it preserves the translation equivariance of the feature map: a shifted spectrum produces shifted but identically weighted features. In a preliminary ablation study, importance gating was found to accelerate convergence and stabilise training by providing the cross-attention mechanism with a pre-filtered, higher-contrast feature map.

Learnable positional embeddings of dimension~128 (Layer~6) are then added element-wise to the normalised feature vectors. The embedding layer is a trainable lookup table of shape $(150 \times 128)$ that assigns a unique 128-dimensional vector to each of the 150~spectral positions. During the forward pass, the integer indices $[0, 1, \ldots, 149]$ are mapped to their corresponding rows in this table, yielding 150~position-specific vectors. These are added component-wise to the feature vectors produced by the convolutional stage, which have the same shape~$(150, 128)$. Without this addition, two identical spectral features located at different wavelengths would produce identical feature vectors, because the convolutional filters are applied uniformly across all positions (translation equivariance). The positional embedding breaks this ambiguity: after the addition, each feature vector encodes both \emph{what} the local spectral morphology looks like (from the convolution) and \emph{where} in the spectrum it is located (from the embedding). The embedding vectors are randomly initialised and learned jointly with all other network parameters during training, so the network determines autonomously how much positional specificity is beneficial, so that each of the 150~spectral positions is represented by a single vector that carries both morphological content from the convolution and spatial context from the embedding. 

These enriched feature vectors serve as \emph{keys} and \emph{values} in the subsequent cross-attention layer (Layer~7). A set of 16~learnable \emph{probe} vectors, each of dimension~128 and shared across all samples in a batch, serve as the corresponding \emph{queries}. Each probe learns to encode a specific spectral question, for example: ``Is there a red-ink absorption peak in this general spectral neighbourhood?'' A multi-head cross-attention layer (8~heads, key dimension~128) computes the interaction between these fixed probes and the position-enriched feature map. Because each feature vector carries both morphological and positional information after the element-wise addition, the attention weights are influenced not only by \emph{what} a local feature looks like but also by \emph{where} in the spectrum it is located. Likewise, the aggregated output inherits approximate spatial context from the values, since the same fused vectors are summed during the weighted aggregation. This enables the model to differentiate between physically distinct chromophores that share similar local morphologies but absorb in different spectral regions, for instance red ink near 550~nm versus black ink, whose absorption extends across the full spectral range or a potential blue ink with the same spectral shape.

The cross-attention output has the shape (batch, 16, 128): the original sequence length of~150 is collapsed by weighted aggregation over all spectral positions (Layer~8 applies the residual connection and layer normalisation). The result encodes \emph{which} morphological features were found and retains an approximate sense of \emph{where} they occurred through the positional component of the aggregated values, but is no longer sensitive to the exact spectral position. This spatial collapse is the fundamental source of shift invariance: a small spectral shift merely redistributes the attention weights to neighbouring positions without substantially altering the aggregated probe activations. Because the positional embeddings are additive and learned jointly with the convolutional features, they act as a soft spatial prior rather than a rigid positional encoding; for small shifts the morphological component dominates the attention scores, so the invariance degrades gradually with increasing shift magnitude rather than breaking abruptly, as confirmed experimentally in section~\ref{sec_robustness}.

A subsequent multi-head self-attention layer (8~heads, Layers~9--10) allows the probes to exchange contextual information with one another (e.g.\ ``Probe~A found an absorber $\rightarrow$ Probe~B should look for scattering''), before a two-layer feed-forward network with an intermediate dimension of~256 and GELU activation (Layers~11--13) provides additional non-linear processing. Residual connections and layer normalisation are applied after each block. The 16~probe vectors are finally flattened into a 2048-dimensional representation (Layer~14).

A fully connected bottleneck layer with 64~neurons and GELU activation (Layer~15) compresses this representation before it enters the latent space. This intermediate step provides an additional non-linear transformation that gradually distils the relevant information, avoiding the abrupt dimensionality reduction from~2048 directly to three latent neurons.

The latent space (Layer~16) consists of three neurons with softplus activation, initialised with a LecunUniform distribution and regularised with an L1 activity penalty of $10^{-4}$. The number of latent neurons corresponds to the number of physical constituents: one scatterer~(intralipid) and two absorbers~(red ink and black ink). Each neuron encodes the relative concentration of exactly one constituent: Neuron~0 encodes intralipid, Neuron~1 red ink, and Neuron~2 black ink. The latent activation vector therefore constitutes a direct, physically interpretable representation of the sample composition, accessible as the \textit{decode\_output} of the model.

\textbf{Physics-constrained decoder.} The decoder is retained in its original design from~\cite{ni24}: a single fully connected layer with 300 neurons, linear activation, no bias, and kernel weights initialised to one (Layer~17). The custom Keras weight constraint implementing the physical decoupling was re-implemented for compatibility with TensorFlow~2.19, as the original code was no longer functional under the updated Keras API; the functional behaviour of the constraint is identical to the original. The 300 output values represent the interleaved, flattened spectral values of $\mu_a(\lambda)$ (even indices) and $\mu_s'(\lambda)$ (odd indices), which are subsequently reshaped into the output array of shape $(150, 2)$ (Layer~18), the \textit{encode\_output} of the model. The kernel weights of the decoder are subject to two hard constraints: (i) all weights are restricted to non-negative values, and (ii) the neuron representing the scatterer~(Neuron~0) may only connect to the scattering output at odd indices, while the neurons representing the absorbers~(Neurons~1 and~2) may only connect to the absorption output at even indices. This constraint enforces a strict physical decoupling of absorption and scattering pathways and guarantees that the decoder operates as a non-negative linear mixture model, in which each latent neuron value corresponds to the relative concentration of a single constituent. This architecture implicitly assumes that each constituent contributes exclusively to either absorption or scattering. In the present phantom system, this assumption is strictly satisfied by intralipid (pure scatterer) and approximately satisfied by the two inks: while both inks are conventionally treated as pure absorbers, Indian ink is a carbon-particle suspension that exhibits measurable scattering in addition to its dominant absorption. The implications of this model misspecification are examined in section~\ref{sec_discussion}.

\begin{table*}[!htb]
	\centering
	\footnotesize\setlength{\tabcolsep}{2pt}
	\begin{tabular}{|c|p{3.8cm}|p{3.8cm}|c|l|}
		\hline
		Layer & Type & Parameters & Activation & Initializer \\
		\hline
		-- & Controllable\newline Augmentation & max\_shift = 7, max\_noise = 0.03, ratio = 0.65 & -- & -- \\
		\hline
		\multicolumn{5}{|c|}{\textit{BiLT Scanner}} \\
		\hline
		1 & Conv1D & filters = 128, kernel size = 23 & gelu & -- \\
		2 & LayerNormalization & -- & -- & -- \\
		3 & MaxPooling1D & pool size = 3, strides = 1 & -- & -- \\
		4 & LayerNormalization & -- & -- & -- \\
		5 & Importance Gating (optional) & statistics: mean/std/max;\newline MLP: $3 \rightarrow 16 \rightarrow 1$ & relu / sigmoid & -- \\
		6 & Positional Embedding & input dim = 150, output dim = 128 & -- & -- \\
		7 & Cross-Attention & 8 heads, key dim = 128; 16 probes & -- & glorot\_uniform \\
		8 & LayerNormalization (+ residual) & -- & -- & -- \\
		9 & Self-Attention & 8 heads, key dim = 128 & -- & -- \\
		10 & LayerNormalization (+ residual) & -- & -- & -- \\
		11 & Dense (FFN) & 256 & gelu & -- \\
		12 & Dense (FFN) & 128 & -- & -- \\
		13 & LayerNormalization (+ residual) & -- & -- & -- \\
		14 & Flatten & $16 \times 128 = 2048$ & -- & -- \\
		\hline
		\multicolumn{5}{|c|}{\textit{Bottleneck and latent space}} \\
		\hline
		15 & Dense & 64 & gelu & -- \\
		16 & Dense (latent) & 3, activity reg.\ L1 = $10^{-4}$ & softplus & LecunUniform \\
		\hline
		\multicolumn{5}{|c|}{\textit{Decoder}} \\
		\hline
		17 & Dense & 300, no bias, kernel constraint & linear & ones \\
		18 & Reshape & $(150, 2)$ & -- & -- \\
		\hline
	\end{tabular}
	\caption{Layers and parameters of the BiLT-Autoencoder. The augmentation layer is active during training only. The importance gating (Layer~5) is optional and can be disabled via a flag. The total number of trainable parameters is approximately 1.3\,million, comparable to RamanNet~\cite{ramannet2023}, a general-purpose transformer for Raman spectra with a similar parameter count, and lightweight by deep-learning standards yet substantially more expressive than classical calibration models such as PLS.}
	\label{tab_layers_nn}
\end{table*}

\subsubsection{Loss functions}

Two separate loss functions are applied to the two model outputs. The reconstruction loss for the spectral output (\textit{encode\_output}) combines three terms:
\begin{equation}
    \mathcal{L}_\mathrm{output} = 20 \cdot \mathrm{MSE}(\mu_a) + \mathrm{MSE}(\mu_s')
    + \alpha\,\mathcal{L}_\mathrm{log} + \beta\,\mathcal{L}_\mathrm{swap}.
    \label{eq_loss_output}
\end{equation}
The first two terms form the baseline weighted MSE. The factor of 20 compensates for the approximately one order of magnitude difference between the absolute values of $\mu_a$ and $\mu_s'$; without this weighting, the dominant MSE contribution from $\mu_s'$ would cause the network to neglect the absorption channel.

The log-scale term $\mathcal{L}_\mathrm{log}$ addresses a systematic bias at low absorber concentrations, where the absolute value of $\mu_a$ is small and the weighted MSE provides insufficient gradient to correct spectral shape errors:
\begin{equation}
    \mathcal{L}_\mathrm{log} = \mathbb{E}\!\left[\left(\log(\mu_a + \varepsilon) - \log(\hat{\mu}_a + \varepsilon)\right)^2\right], \quad \varepsilon = 10^{-6},
    \label{eq_loss_log}
\end{equation}
where $\hat{\mu}_a$ denotes the predicted value. On a logarithmic scale, a given relative error receives equal weight regardless of the absolute concentration level. The coefficient is set to $\alpha = 0.1$ to keep the log-term as a soft correction without dominating the main MSE gradient.

The swap penalty $\mathcal{L}_\mathrm{swap}$ targets a specific failure mode in which the model assigns an erroneously high absorption amplitude while simultaneously underestimating scattering, a confusion that arises for broadband absorbers whose spectrally flat signature resembles the smooth power-law decrease of scattering:
\begin{equation}
    \mathcal{L}_\mathrm{swap} = \mathbb{E}\!\left[
        \mathrm{ReLU}\!\left(\bar{\mu}_a^\mathrm{pred} - \bar{\mu}_a^\mathrm{true}\right)
        \cdot
        \mathrm{ReLU}\!\left(\bar{\mu}_s^{\prime\,\mathrm{true}} - \bar{\mu}_s^{\prime\,\mathrm{pred}}\right)
    \right],
    \label{eq_loss_swap}
\end{equation}
where $\bar{\cdot}$ denotes the spectral mean over all 150 wavelength points, computed per spectrum. The product structure ensures that the penalty is active only when both conditions hold simultaneously for the same sample: absorption is overestimated \emph{and} scattering is underestimated. This avoids penalising samples where either channel deviates in isolation, and is robust to spectrally selective absorbers (e.g.\ red ink) whose narrow absorption peak raises the local $\mu_a$ above $\mu_s'$ but leaves the spectral mean well below. The coefficient is set to $\beta = 1.0$.

The latent loss applied to the \textit{decode\_output} combines two physically motivated penalty terms:
\begin{equation}
    \mathcal{L}_\mathrm{latent} = \underbrace{\mathbb{E}\!\left[\,|\min(0,\,\mathbf{z})|\,\right]}_{\mathcal{L}_\mathrm{neg}}
    \;+\; \lambda_s \underbrace{\mathbb{E}\!\left[z_1 \cdot z_2\right]}_{\mathcal{L}_\mathrm{excl}},
    \label{eq_loss_latent}
\end{equation}
where $\mathbf{z}$ denotes the latent activation vector. The first term, $\mathcal{L}_\mathrm{neg}$, penalises negative latent values, reinforcing the physical constraint that relative concentrations must be non-negative. The second term, $\mathcal{L}_\mathrm{excl}$, is the mutual exclusivity penalty: it penalises the simultaneous activation of both absorber neurons ($z_1$ and $z_2$), discouraging the network from splitting the representation of a single absorber across two neurons, a systematic error observed in the preceding study~\cite{ni24}. The weighting factor is set to $\lambda_s = 0.1$.

\subsubsection{Training procedure}
\label{sec_training}

All computations were performed in Python using TensorFlow~\cite{tensorflow2015_whitepaper} and the model was trained on an NVIDIA RTX~3090 GPU. The Adam optimiser~\cite{kingma2014adam} was used with a global gradient clip norm of 1.0 and a batch size of 64. To achieve stable convergence in the presence of strong data augmentation, training is organised into three consecutive phases, summarised in table~\ref{tab_training}.

\textbf{Phase~1 -- Clean training} (1000 epochs): The model is trained without any augmentation (strength = 0) using a cosine learning rate decay from $10^{-3}$ to $2 \times 10^{-5}$. This phase allows the network to first establish a reliable mapping between undistorted spectra and OP.

\textbf{Phase~2 -- Augmentation ramp-up} (4000 epochs): The augmentation strength is linearly increased from 0 to 1 over the course of the phase, while the learning rate is held constant at $2 \times 10^{-5}$. Gradually introducing the augmentation, rather than applying it at full strength from the outset, was found to be essential for stable convergence; applying the full augmentation from the beginning destabilised training in preliminary experiments.

\textbf{Phase~3 -- Cooldown} (3000 epochs): Training continues with full augmentation and a smooth cosine annealing of the learning rate from $2\times10^{-5}$ to $2\times10^{-6}$, continuing seamlessly from Phase~2 without a learning rate jump. This avoids displacing the model from the well-converged minimum found at the end of Phase~2. Early stopping with a patience of 1500 epochs monitors the validation loss from epoch~4500 onwards and restores the best weights found during training.

During training, the augmentation layer applies two types of perturbation to randomly selected samples (65\% probability per sample): (i) a sub-pixel spectral shift between $-7$ and $+7$ spectral bands, implemented via linear interpolation on a symmetrically padded signal to avoid boundary artefacts; and (ii) additive Gaussian noise with a standard deviation drawn uniformly from $[0,\, 3\%]$. The validation set consists of a 50/50 mixture of clean test spectra and augmented copies of those same spectra, ensuring that the validation metric captures both in-distribution accuracy and robustness to perturbation simultaneously.

\begin{table*}[!htb]
    \centering
    \footnotesize
    \begin{tabular}{|l|c|p{3.8cm}|p{1.8cm}|p{2.5cm}|}
        \hline
        Phase & Epochs & Learning rate & Aug.\ strength & Early stopping \\
        \hline
        1 (Clean)    & 1000 & Cos.\ decay: $10^{-3} \to 2{\times}10^{-5}$              & off           & --                        \\
        2 (Ramp-up)  & 4000 & Const.: $2{\times}10^{-5}$                                & lin.\ $0\to1$ & --                        \\
        3 (Cooldown) & 3000 & Cos.\ anneal.: $2{\times}10^{-5} \to 2{\times}10^{-6}$   & full          & pat.\ 1500 (ep.\ 4500)    \\
        \hline
    \end{tabular}
    \caption{Three-phase curriculum training strategy. All phases use the Adam optimiser with global gradient clip norm of 1.0 and a batch size of 64.}
    \label{tab_training}
\end{table*}

\section{Results and discussion}

\subsection{Training convergence}

\begin{figure}[!htb]
    \centering
    \includegraphics[width=0.55\textwidth]{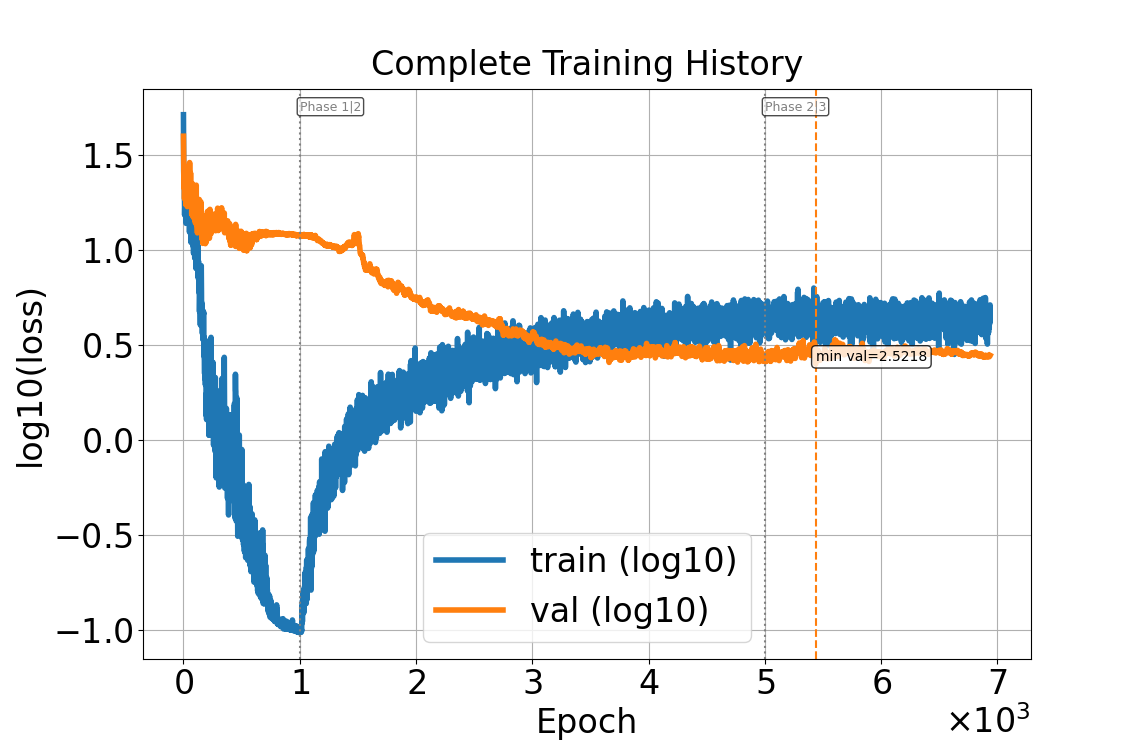}
    \caption{Complete training history of the BiLT-Autoencoder. Training loss (blue) and validation loss (orange) are shown on a logarithmic scale. The three curriculum phases are reflected in the loss curve: steep descent during Phase~1 (clean training), a shoulder during Phase~2 (augmentation ramp-up), and fine convergence during Phase~3 (cooldown). The dashed vertical line marks the early-stopping trigger point.}
    \label{im_training}
\end{figure}

Figure~\ref{im_training} shows the complete training history of the BiLT-Autoencoder over all 8\,000 epochs. The three-phase curriculum structure is clearly reflected in the loss curve. During Phase~1 (clean training, epochs 1--1000), the training loss drops steeply as the model rapidly establishes an accurate mapping from undistorted spectra to OP. The transition to Phase~2 (augmentation ramp-up) introduces a characteristic shoulder in both curves as the model must simultaneously maintain accuracy on clean spectra and adapt to progressively stronger spectral perturbations. Throughout Phase~2 and Phase~3, the validation loss closely tracks the training loss. This close agreement confirms that the mixed validation set, comprising equal parts clean and augmented spectra, accurately captures the model's generalisation behaviour and that the model does not overfit despite the relatively modest dataset size. Early stopping triggers at approximately epoch~4500 and restores the best checkpoint found during the cooldown phase.

\subsection{Prediction accuracy}

Table~\ref{tab_results} summarises the predictive performance of the BiLT-Autoencoder across all evaluated conditions, reporting both the coefficient of determination~(R²) and the mean absolute percentage error~(MAPE) for each output channel. MAPE is computed only over wavelength points where the ground-truth value exceeds 1\% of its spectral maximum, to exclude near-zero regions where relative errors are not physically meaningful.

\begin{table*}[!htb]
    \centering
    \footnotesize
    \begin{tabular}{|l|c|c|c|c|}
        \hline
        Dataset & R²\;$\mu_a$ & MAPE\;$\mu_a$ & R²\;$\mu_s'$ & MAPE\;$\mu_s'$ \\
        \hline
        Training (clean)                          & 0.991 & 9.3\%  & 0.991 & 9.0\%  \\
        Test (clean)                              & 0.979 & 16.2\% & 0.975 & 9.7\%  \\
        Test (noisy/shifted, mixed validation)    & 0.917 & 25.3\% & 0.917 & 19.9\% \\
        \hline
        Sim.\ spectrometer (no blur)               & 0.975 & 16.8\% & 0.976 & 9.7\%  \\
        Sim.\ spectrometer (FWHM $\approx 4.7$\,nm)  & 0.972 & 17.0\% & 0.974 & 9.5\%  \\
        Sim.\ spectrometer (FWHM $\approx 14.1$\,nm) & 0.971 & 17.1\% & 0.975 & 9.9\%  \\
        Sim.\ spectrometer (FWHM $\approx 23.6$\,nm) & 0.960 & 17.6\% & 0.974 & 10.2\% \\
        \hline
    \end{tabular}
    \caption{Predictive performance of the BiLT-Autoencoder under all evaluated conditions. The simulated spectrometer rows correspond to test spectra processed in the following order: (1)~optional Gaussian blur on the full 150-point spectrum to simulate a broader instrument line shape, (2)~downsampling to 60 points, (3)~random shift of $\pm1$ band, (4)~1\% additive noise, (5)~upsampling back to 150 points. Applying the blur before downsampling and noise reflects the physical measurement process, in which the instrument line function acts on the optical signal prior to detection. The model was not retrained or recalibrated for any of these conditions. Note that the high robustness to increasing FWHM is partly a consequence of the spectrally broad features of the phantom constituents used in this study; the test is a necessary but not sufficient condition for generalisation to samples with narrower absorption features.}
    \label{tab_results}
\end{table*}

On clean, unperturbed spectra the model achieves R²\,$= 0.979$ ($\mu_a$) and R²\,$= 0.975$ ($\mu_s'$) on the held-out test set, with MAPE values of 16.2\% and 9.7\% respectively. These values are comparable to those reported for the preceding dense mixed autoencoder~\cite{ni2025automated}, confirming that replacing the fully connected encoder with the BiLT scanner does not compromise in-distribution predictive accuracy. The slightly larger MAPE for $\mu_a$ reflects the inherent difficulty of accurately recovering low-concentration absorption spectra, whose absolute values span more than one order of magnitude across the sample set. The negligible gap between training and test R² supports the absence of overfitting.

\subsection{Robustness to spectral perturbations}\label{sec_robustness}

\begin{figure}[!htb]
	\centering
	\includegraphics[width=0.65\textwidth]{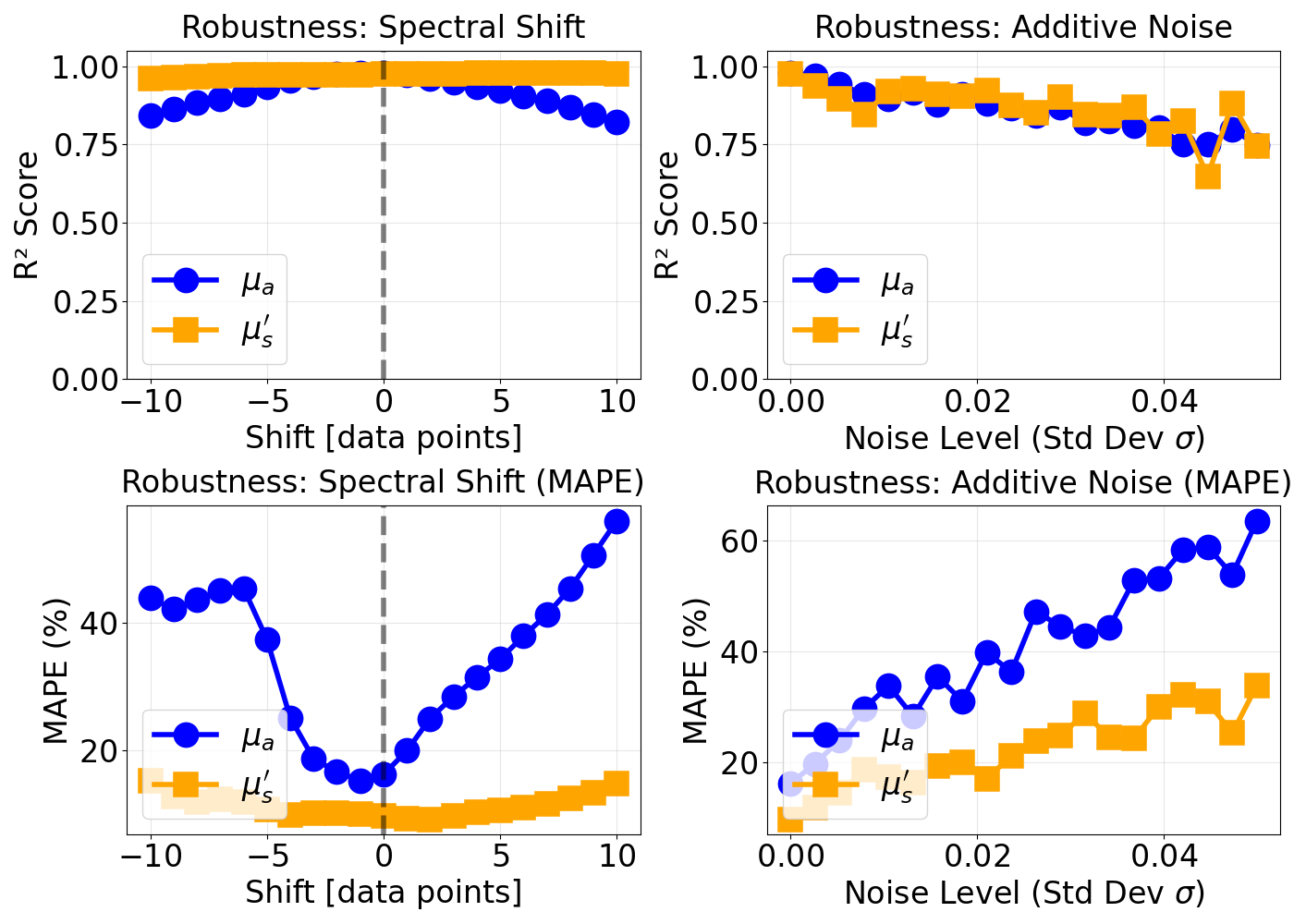}
	\caption{Robustness of the BiLT-Autoencoder to spectral perturbations. Top row: R² scores for $\mu_a$ (blue circles) and $\mu_s'$ (orange squares) as a function of applied spectral shift~(left) and additive Gaussian noise standard deviation~(right). Bottom row: corresponding MAPE values. Shift robustness is intrinsic to the cross-attention architecture; noise robustness requires augmentation training (see text).}
	\label{im_robustness}
\end{figure}

Figure~\ref{im_robustness} presents the R² scores and MAPE of both output channels as a function of applied spectral shift~(top row) and additive Gaussian noise~(bottom row).

Under spectral shift, $\mu_s'$ maintains near-perfect predictive accuracy (R²~$\approx$~0.99) across the entire tested range of $\pm10$ spectral bands, with negligible degradation even at the most extreme offsets. The absorption coefficient $\mu_a$ likewise remains highly accurate within the training augmentation range of $\pm7$ bands (R²~$>$~0.90), with a steeper but gradual decline at larger offsets. The difference in shift tolerance between the two channels reflects the spectral character of the respective signals: the reduced scattering spectrum is a smooth, slowly varying power-law function whose overall morphology is largely preserved under a rigid wavelength offset, whereas the absorption spectrum contains sharper spectral features that are partially displaced outside the measured window at large shifts. The model generalises well beyond its training augmentation range, demonstrating that the cross-attention mechanism has learned genuine morphological representations rather than memorised position-specific features.

Under additive Gaussian noise, both channels exhibit a smooth and monotonic decrease in R² as the noise standard deviation increases from 0 to 5\%. The degradation is gradual and predictable, with no threshold or catastrophic failure behaviour. This smooth degradation reflects the inherent averaging effect of the cross-attention mechanism across multiple spectral positions, which suppresses uncorrelated point-wise noise more effectively than a dense encoder that weights all positions simultaneously and rigidly.

The MAPE curves complement the R² analysis by providing a scale-independent error measure that is particularly informative for $\mu_a$, whose dynamic range spans more than one order of magnitude across samples. At zero shift and zero noise, MAPE reaches 16.2\% for $\mu_a$ and 9.7\% for $\mu_s'$ (table~\ref{tab_results}), and degrades gracefully under increasing perturbation.

To disentangle the contributions of architecture and training procedure, an ablation experiment was conducted in which the model was trained without any shift augmentation (ramp-up disabled, shift strength fixed at zero throughout training). The resulting model retained its shift robustness: R² scores under spectral shift were comparable to the fully augmented model, confirming that shift invariance is an intrinsic property of the cross-attention architecture and does not depend on shift augmentation during training. Noise robustness, by contrast, depended critically on the augmentation ramp-up: the ablated model degraded rapidly under additive noise, with R² collapsing at noise levels well below $\sigma = 1\%$. These results establish a clear separation of mechanisms: shift invariance arises from morphological feature extraction via cross-attention, while noise robustness is conferred by the curriculum augmentation strategy.
Taken together, these results demonstrate that the BiLT-Autoencoder achieves the intended instrument-independence: a model trained with augmentation generalises to the full $\pm10$ band range with only moderate accuracy loss, without any post-hoc calibration transfer or domain adaptation. This stands in direct contrast to the expected behaviour of a dense encoder, which would degrade catastrophically under even small wavelength offsets due to its strict dependence on absolute pixel positions~\cite{chatzidakis2019calibration}.
While architectural shift invariance through global average pooling has been demonstrated for spectral classification~\cite{chatzidakis2019calibration}, complete shift invariance is neither achievable nor desirable for quantitative spectral reconstruction. Classification requires only a discrete label and tolerates complete loss of spatial structure; global average pooling achieves invariance precisely by discarding all positional information. Spectral reconstruction, by contrast, requires that positional information be preserved -- a globally shift-invariant encoder would be unable to map features back to wavelength positions. The appropriate target is \emph{local} shift invariance: robustness to the small calibration drifts encountered in practice, while retaining enough spectral structure for accurate reconstruction beyond the invariant range. Cross-attention achieves this balance: probes aggregate content-based features that are robust to small shifts, while the linear decoder retains the spectral resolution needed for quantitative output. The graceful degradation observed beyond $\pm$10 bands reflects this design -- a pure global-pooling encoder would show no degradation at any shift, but would also be unable to reconstruct spectra at all.

\subsection{Spectral reconstruction and attention maps}

\begin{figure}[!htb]
    \centering
    \includegraphics[width=0.95\textwidth]{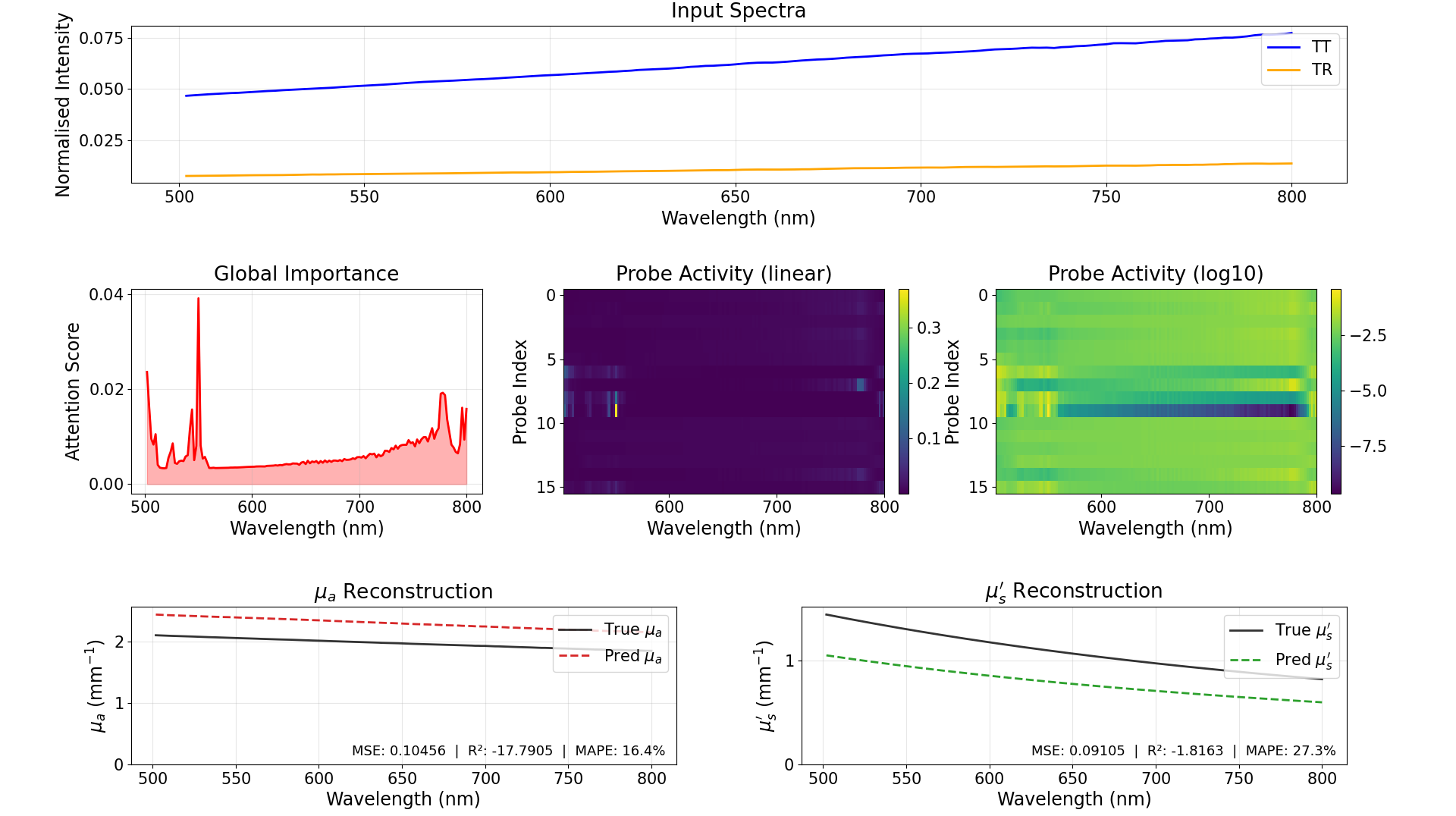}
    \caption{Analysis of a clean broadband absorber sample (black ink). The model
    identifies discrete spectral landmarks at 500--530\,nm and deploys a single
    dedicated probe at ${\approx}780$\,nm, where lower absorption and scattering
    yield higher transmittance and better SNR. The strongly negative R$^2$ for both
    output channels is a known artefact of spectrally flat targets; MAPE is the
    appropriate accuracy metric.}
    \label{im_results_black_clean}
\end{figure}

\begin{figure}[!htb]
    \centering
    \includegraphics[width=0.95\textwidth]{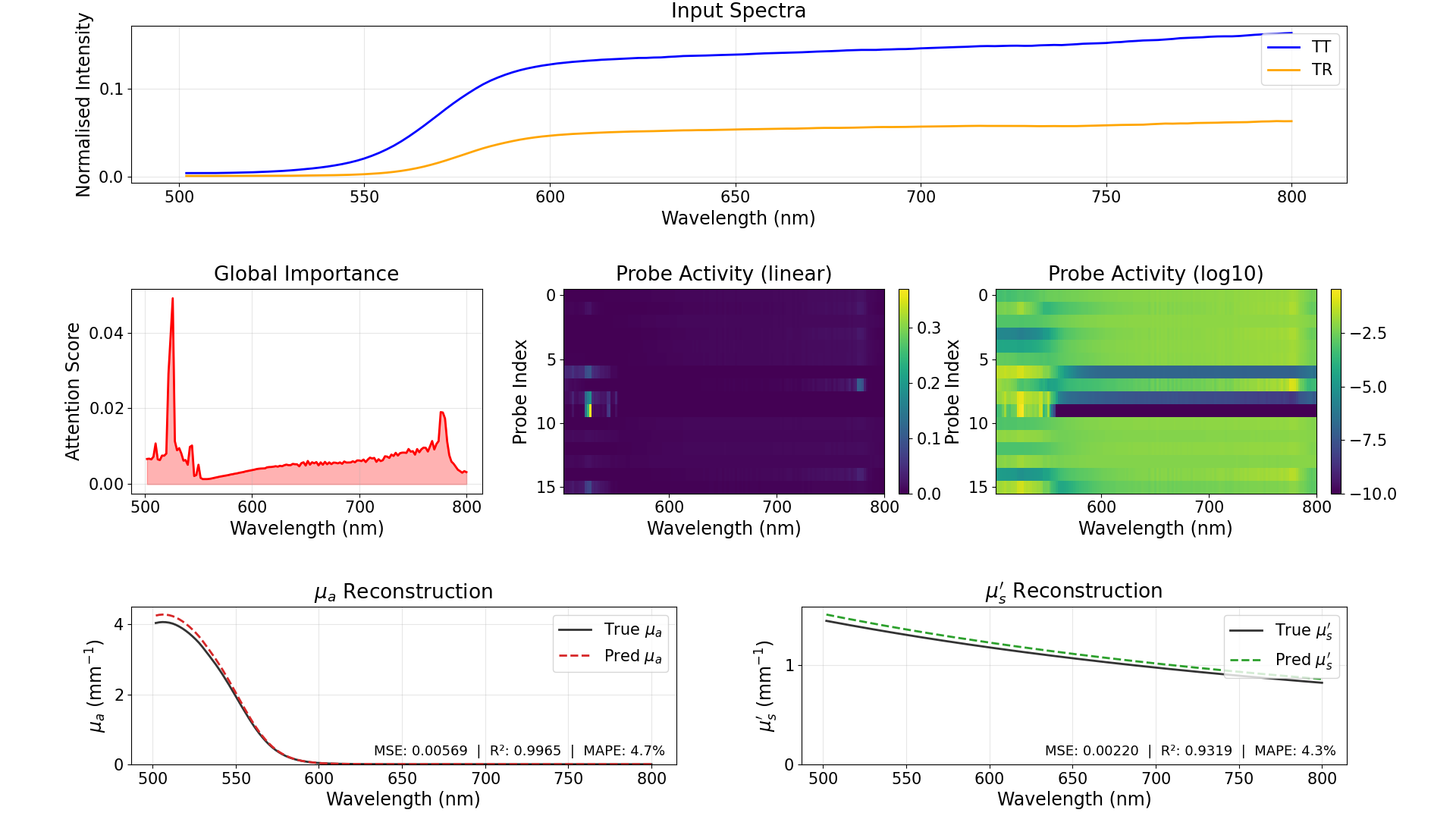}
    \caption{Analysis of a clean spectrally selective absorber (red ink). The
    attention structure mirrors the black ink case: discrete spikes across the
    absorption edge (500--550\,nm) plus a dedicated probe at ${\approx}780$\,nm.
    The broader short-wavelength coverage compared to black ink reflects the
    spectrally resolved absorption onset of red ink.}
    \label{im_results_red_clean}
\end{figure}

\begin{figure}[!htb]
    \centering
    \includegraphics[width=0.95\textwidth]{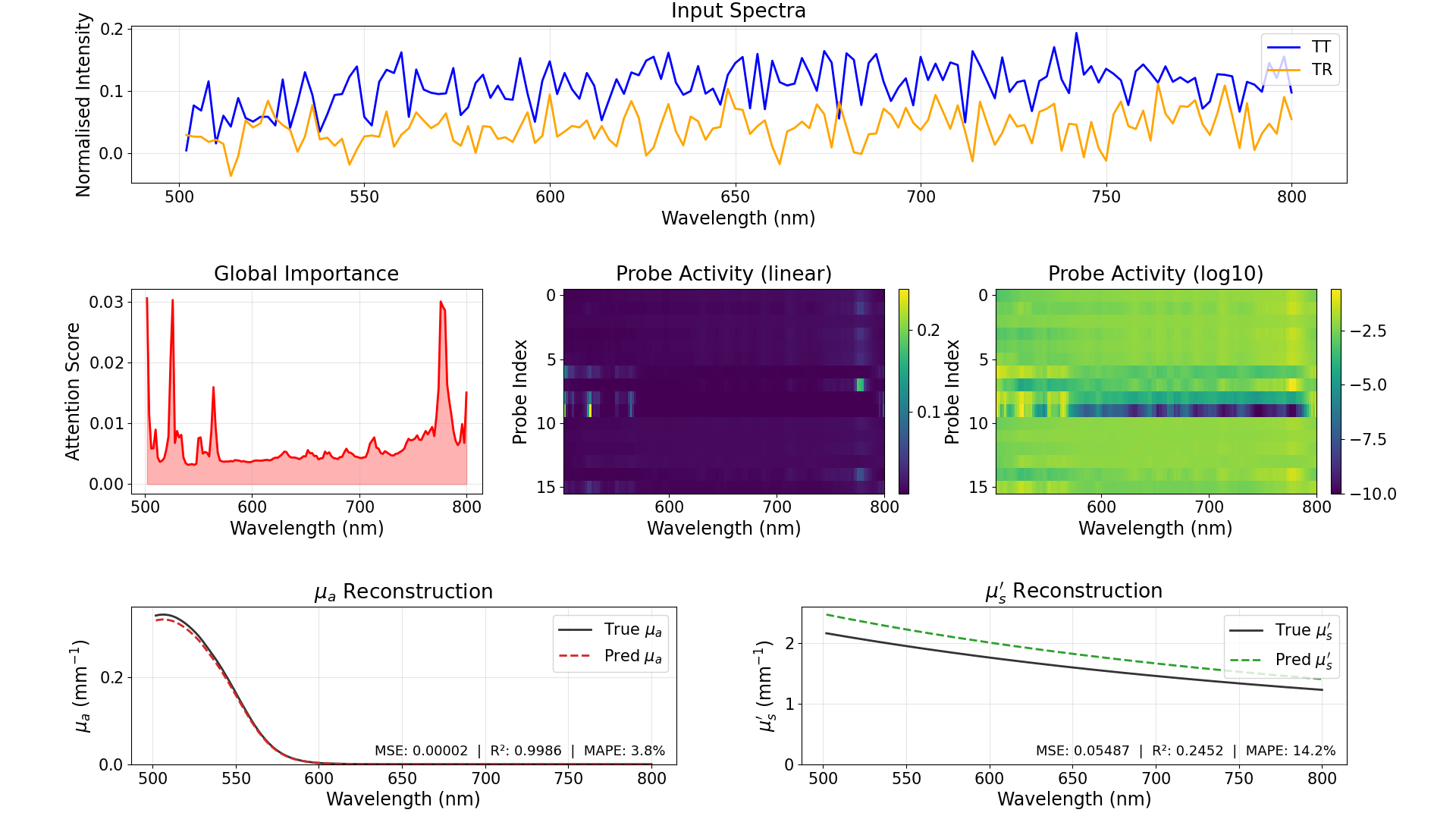}
    \caption{Analysis of a severely noise-corrupted spectrally selective absorber.
    Under noise, an ensemble of probes converges on ${\approx}780$\,nm (highest SNR),
    greatly amplifying the global attention there compared to the clean case, while
    the short-wavelength spikes remain narrow. The $\mu_a$ reconstruction is highly
    accurate (R$^2 = 0.999$), while $\mu_s'$ is more strongly affected
    (R$^2 = 0.245$), reflecting the weaker morphological signal of the slowly
    varying scattering spectrum.}
    \label{im_results_red_noisy}
\end{figure}

To evaluate both the quantitative fidelity and the internal reasoning of the BiLT-Autoencoder, consolidated analyses are presented for three representative samples. Figures~\ref{im_results_black_clean} to \ref{im_results_red_noisy} display the input spectra (top), the global spectral attention and per-probe activity (middle), and the resulting spectrally resolved reconstructions for $\mu_a$ and $\mu_s'$ (bottom). The global importance curves (mean attention across all heads and probes) and the per-probe heatmaps provide a direct window into the morphological features the model relies upon to determine constituent concentrations.

\textbf{Broadband absorber, clean (Black ink, Figure~\ref{im_results_black_clean}).} Black ink absorbs uniformly across the visible spectrum, producing slowly varying, relatively flat input signals with no distinct morphological features. The global importance curve is not smooth but consists of discrete, sharp spikes concentrated in the 500--530\,nm range, indicating that the model identifies subtle fine-scale structures in the otherwise featureless signal as anchor points. Beyond these discrete landmarks, a broad, continuously distributed background attention is present across the intermediate and long-wavelength region (${\gtrsim}600$\,nm) and rises monotonically toward longer wavelengths. When integrated over wavelength, this background component carries more total attention weight than the discrete short-wavelength spikes, reflecting a strong, diffuse preference for the long-wavelength end of the spectrum that arises from a large number of non-specialised probes attending collectively to the full spectrum with a near-identical, SNR-weighted attention distribution. In addition, a single dedicated probe is anchored at the high-transmittance long-wavelength end of the spectrum (${\approx}780$\,nm), riding on top of this background as the most prominent long-wavelength anchor. Both the background and the dedicated probe are physically motivated by the same principle: lower absorption and scattering coefficients at longer wavelengths yield higher transmitted intensity and therefore a better signal-to-noise ratio, making the long-wavelength region a more reliable amplitude reference. In the per-probe activity plot (linear scale), the 780\,nm probe stands out as the most prominent single anchor; the non-specialised probe majority is nearly invisible at this scale, yet collectively produces the dominant background component visible in the global importance curve. The resulting $\mu_a$ and $\mu_s'$ reconstructions capture the correct spectral morphology but exhibit moderate amplitude offsets. These offsets are responsible for strongly negative R$^2$ values for both channels: a well-known artefact of spectrally flat targets where the total sum of squares is extremely small, so even a modest constant offset drives R$^2$ deeply negative. MAPE is the physically meaningful accuracy metric in this regime.

\textbf{Spectrally selective absorber, clean (Red ink, Figure~\ref{im_results_red_clean}).} Red ink exhibits strong absorption at short wavelengths and high transmission at longer wavelengths, producing a characteristic inflection curve. The attention structure is fundamentally the same as for black ink: discrete short-wavelength spikes, a broad rising background at longer wavelengths, and a dedicated probe at ${\approx}780$\,nm. The key difference lies in the extent of the short-wavelength activity: for red ink the spikes span the full absorption edge from 500 to ${\approx}550$\,nm, tracking the spectrally resolved onset of the absorption band, whereas for black ink the activity is concentrated in the narrower 500--530\,nm window. This broader interrogation range reflects the distinct spectral character of the absorber and the richer morphological information available at the absorption edge. The $\mu_a$ reconstruction accurately captures both the spectral shape and the absolute amplitude (R$^2 = 0.997$, MAPE$\,= 4.7\%$), and $\mu_s'$ is similarly well recovered (R$^2 = 0.932$, MAPE$\,= 8.3\%$).

\textbf{Spectrally selective absorber with severe noise (Red ink, Figure~\ref{im_results_red_noisy}).} Under heavy additive noise the input spectrum is severely corrupted, masking the smooth inflection features. The basic attention structure mirrors Figure~\ref{im_results_red_clean}: discrete spikes at 500--550\,nm, a broad rising background at longer wavelengths, and concentrated activity at ${\approx}780$\,nm. The critical difference is quantitative: under noise, an entire ensemble of probes converges on the 780\,nm region rather than the single probe observed under clean conditions, substantially amplifying both the global attention at that wavelength and the already dominant long-wavelength background. More probes are recruited in total compared to the clean case, consistent with an implicit averaging strategy: by pooling information across multiple probes at the highest-SNR spectral position, the model suppresses uncorrelated point-wise noise artefacts without broadening its individual attention peaks. The primary short-wavelength spikes remain narrow. The $\mu_a$ reconstruction maintains high accuracy despite the severe signal degradation (R$^2 = 0.999$, MAPE$\,= 3.8\%$). The $\mu_s'$ prediction is more strongly affected by the noise (R$^2 = 0.245$, MAPE$\,= 14.2\%$): the slowly varying power-law shape of the scattering spectrum offers weaker morphological signal than the sharply defined absorption edge, making it more susceptible to amplitude errors under noise.

\subsection{Discussion}\label{sec_discussion}

The attention maps presented above reveal a consistent two-component probe strategy across all three cases: a small number of specialised probes concentrate sharply at feature-relevant wavelengths, while the non-specialised majority collectively produces the monotonically rising, SNR-weighted background. The specialised probes are functionally analogous to narrow bandpass filters in classical spectroscopic instrumentation, learned entirely from data without any explicit spectral prior. The most significant finding is the model's adaptive response to noise: rather than broadening its attention or falling back to low-resolution tracking, the network recruits an ensemble of probes at the high-SNR long-wavelength extreme, achieving implicit spectral averaging while keeping individual attention peaks sharp. This dynamic reallocation is structurally impossible for a conventional dense encoder, whose contribution weights are fixed at inference time regardless of local noise conditions or spectral translations. The progressive exhaustion of reliable anchor regions under increasing noise provides a mechanistic explanation for the smooth, monotonic R$^2$ degradation observed in figure~\ref{im_robustness}: the model degrades gracefully because it depletes its pool of high-SNR anchors gradually rather than failing at a discrete threshold. Together, these observations establish the attention maps as a genuine window into the model's internal reasoning, confirming that the BiLT-Autoencoder operates via physically grounded morphological feature extraction rather than point-wise memorisation.

A further observation contextualises the difficulty of the present problem relative to other machine-learning approaches in analytical spectroscopy. In many common spectroscopic modalities, both the measured signal and the target quantity are spectra in the same physical domain, connected by a forward model that is linear or nearly so. In fluorescence spectroscopy, for example, the emission signal scales proportionally with the concentration of each fluorophore; a network solving the inverse problem need only assign spectral peaks to their sources, since the mapping from constituent concentration to measured intensity is trivially invertible. Similar linearity holds for Raman scattering and for transmission measurements under the Beer--Lambert regime. In the present case, by contrast, the integrating sphere signals encode optical properties through the nonlinear, multiply-scattering radiative transfer equation. The network must therefore invert a cross-domain forward model with no linear analogue, placing the problem in a categorically harder class than spectral peak assignment. A further distinction from supervised spectroscopic regression approaches, such as RamanFormer~\cite{koyun2024ramanformer}, which requires known constituent concentrations as training labels, is that the mixed autoencoder framework identifies and separates constituent spectra in an entirely unsupervised fashion: no concentration labels are required at training time. The constituent identities and spectra emerge purely from the structural constraints imposed by the physics-constrained decoder, making the approach applicable to systems where reference concentrations are unavailable. The high predictive accuracy ($R^2 > 0.97$ on clean test data, degrading gracefully to $R^2 \approx 0.92$ under severe combined noise and spectral shift) and the emergent physically interpretable attention strategies reported here demonstrate that the BiLT architecture is capable of solving this harder class of inverse problem, further strengthening the case for its applicability beyond the turbid-medium domain.

This unsupervised capability addresses a practical bottleneck that is pervasive in biophotonic spectroscopy. In many in vivo and ex vivo settings, the precise chromophore composition of the sample is not known a priori: biological tissue contains a variable, patient-specific mixture of haemoglobin derivatives, melanin, lipids, water, and potentially exogenous agents, whose identities and relative abundances cannot be established by independent analytical means without destroying the sample. Analogous situations arise in Raman spectroscopy and laser-induced breakdown spectroscopy~(LIBS), where the spectral signatures of chemically complex matrices overlap to a degree that makes classical endmember extraction (e.g.\ vertex component analysis or non-negative matrix factorisation) unreliable, particularly when the forward model is nonlinear. The BiLT-Autoencoder offers a path towards resolving this challenge: by training on a sufficiently large and diverse set of tissue measurements, the physics-constrained decoder can, in principle, identify the constituent spectra and their concentrations without any prior knowledge of the sample composition, provided the number of latent neurons is chosen to match the chemical complexity of the system. Demonstrating this capability on real tissue data remains a key objective of ongoing work. A preliminary extension of the present architecture to an overcomplete latent space, in which more neurons are provided than there are known constituents, confirmed that the combined sparsity constraints (L1 activity regularisation together with a spectral-similarity-weighted co-activation penalty on the decoder weights) successfully suppress redundant neurons: in a six-neuron variant trained on the same phantom dataset, three absorber neurons converged to zero activation across all test samples, while the remaining two recovered the correct constituent spectra without prior knowledge of the number of absorbers. A full investigation of this self-determining variant is beyond the scope of the present paper and will be reported separately.

A further point concerns the strict absorber/scatterer decoupling enforced by the decoder constraint. As noted in the methods section, the architecture assumes that each constituent contributes exclusively to either absorption or scattering. Indian ink, however, is a carbon-particle suspension whose scattering cross-section is non-negligible, so this assumption is not strictly satisfied. Despite this model misspecification, the reconstruction accuracy for both $\mu_a(\lambda)$ and $\mu_s'(\lambda)$ remains high across all test conditions. The most likely explanation is that, at the concentrations used in the present phantom system, the scattering contribution of Indian ink is small compared to the dominant scattering of intralipid. The non-negative linear mixture constraint forces a best-fit decomposition: because the decoder can only attribute scattering to the intralipid neuron, any ink-induced scattering is absorbed into the intralipid term as a minor perturbation, while the ink neuron captures the dominant absorption component. This implicit redistribution produces only a small systematic bias rather than a qualitative failure, precisely because the misspecified contribution is weak relative to the correctly specified one. From an application perspective, this robustness to moderate model misspecification is encouraging: in biological tissue, chromophores such as melanin exhibit both absorption and scattering, and haemoglobin-containing erythrocytes scatter appreciably in addition to absorbing. A model that tolerates minor violations of the pure-constituent assumption is therefore better positioned for translation to realistic samples than one that requires strict compliance.

\section{Conclusion}

This work has presented the BiLT-Autoencoder, a neural network framework for the simultaneous, constituent-level extraction of absorption and scattering spectra from integrating sphere measurements of turbid media. The central architectural innovation is the replacement of the dense encoder used in the preceding mixed autoencoder~\cite{ni2025automated} with a cross-attention scanner inspired by the Byte Latent Transformer~\cite{pagnoni2024byte}, which grants the model intrinsic local shift invariance, the architectural prerequisite for instrument-independent spectral unmixing. The qualifier \emph{local} is deliberate: complete shift invariance -- achievable, for instance, via global average pooling~\cite{chatzidakis2019calibration} -- is not a suitable design target for spectral reconstruction, because it requires discarding all positional information. A reconstruction task demands that spectral structure be preserved; the appropriate goal is therefore robustness to the small calibration drifts encountered in practice, not invariance to arbitrarily large shifts. Cross-attention achieves this balance by aggregating content-based features that are insensitive to small positional offsets while retaining the spectral resolution required by the linear decoder.

The model achieves R²\,$= 0.991$ ($\mu_a$) and R²\,$= 0.991$ ($\mu_s'$) on the training set and R²\,$= 0.979$ / $0.975$ on the held-out test set on clean, unperturbed spectra, matching the predictive accuracy of the fully connected baseline while adding robustness at no cost to in-distribution performance. Under spectral shift, $\mu_s'(\lambda)$ is predicted with R²~$\approx$~0.99 across the full tested range of $\pm10$ spectral bands, and $\mu_a(\lambda)$ remains above R²~$>$~0.90 throughout the $\pm7$-band training augmentation range, with graceful degradation beyond it. Under additive Gaussian noise, both channels degrade smoothly without catastrophic failure. The model also generalises to a simulated alternative spectrometer with a broader instrument line shape (Gaussian blur up to $\sigma = 5$\,px, ${\approx}24$\,nm FWHM), retaining R²~$\approx 0.96$ for $\mu_a$ and R²~$= 0.974$ for $\mu_s'$ without any retraining or recalibration. It should be noted, however, that this result is of limited generality: the absorption features of the phantom constituents used here (broadband black ink, slowly varying red-ink spectrum) are intrinsically wide relative to the instrument bandwidth, so spectral blurring leaves the morphologically relevant features largely intact. Validation on samples with spectrally narrower absorption features would be required to assess FWHM robustness in the general case.

The attention maps confirm that these quantitative gains are grounded in physically meaningful representations: the model autonomously decomposes spectral unmixing into two distinct sub-tasks (morphological alignment for shift estimation and amplitude readout for concentration extraction), which it executes jointly under clean conditions and spatially decouples under severe noise, deploying an ensemble of probes at the high-SNR long-wavelength extreme. Together with the per-probe functional specialisation, these properties establish the BiLT-Autoencoder as an interpretable model in the strict sense: the attention maps are a genuine window into the model's reasoning, not a post-hoc approximation.

The present study is validated on a controlled liquid phantom system using intralipid and two inks as absorbers, and the dataset is relatively modest in size~(496 samples). Extension to real biological tissue or pharmaceutical matrices, which contain a larger number of chemically diverse constituents and are subject to sample-to-sample structural variability, will require larger and more diverse training datasets. Furthermore, while the BiLT-Autoencoder is architecturally robust to wavelength offsets by design, experimental validation of instrument-to-instrument transfer on physical hardware remains an important next step. A further limitation concerns spectral resolution: the absorption and scattering features of the present phantom system are intrinsically broad relative to the instrument bandwidth, making the model's robustness to changes in spectrometer FWHM difficult to assess from this dataset alone. Validation on samples with spectrally narrower absorption features would be required to characterise FWHM robustness in the general case. Finally, the strict absorber/scatterer decoupling enforced by the decoder is an idealisation: Indian ink, one of the two absorbers used here, exhibits non-negligible scattering. As discussed above, the model tolerates this misspecification at the present concentration ratios, but the limits of this tolerance for constituents with stronger scattering contributions remain to be established.

Nevertheless, the results presented here demonstrate that the combination of cross-attention encoding, physics-constrained decoding, and curriculum augmentation provides a principled and practically effective route towards instrument-independent, interpretable spectral unmixing of turbid media. Notably, integrating sphere inversion is a categorically harder problem than the linear, same-domain unmixing tasks common in Raman, fluorescence, or infrared spectroscopy, since it requires inverting a nonlinear, cross-domain forward model governed by the radiative transfer equation. To the author's knowledge, this study represents the first successful adaptation of the Byte Latent Transformer architecture from natural language processing to optical spectroscopy. Given that the BiLT framework achieves high accuracy and interpretability on this harder class of inverse problem, it is anticipated to be well-positioned for broad application across analytical modalities.

\section*{Data availability}
The data is available upon reasonable request.

\section*{Acknowledgements}

The authors gratefully acknowledge funding of the Erlangen Graduate School in Advanced Optical Technologies (SAOT) by the Bavarian State Ministry for Science and Art.

The authors would like to thank the German Research Foundation (DFG-Deutsche Forschungsgemeinschaft) for its support. This work was achieved in the context of the DFG-project "Hyperspektrale Tiefenrekonstruktion durch einen Iso-Punkt-Ansatz zum Verst\"andnis der Licht-Materie-Interaktion in tr\"uben Medien auf einer makroskopischen Skala." (project number 503621669).

\section*{Declaration of Generative AI and AI-Assisted Technologies}
During the preparation of this work, the authors used Claude Sonnet 4.6 (Anthropic) to assist with language editing and revision of the manuscript, including figure captions and discussion sections. After utilising this tool, the authors thoroughly reviewed and edited all content as necessary, and they take full responsibility for the final version of the manuscript.

\bibliographystyle{elsarticle-num}
\bibliography{References}

\end{document}